\pdfoutput=1
\documentclass[aps,prd,twocolumn,superscriptaddress,showpacs,preprintnumbers]{revtex4-1}
\usepackage[colorlinks=true, pdfstartview=FitV, linkcolor=red, citecolor=blue, urlcolor=black, pdftitle={},pdfauthor={Tomoya Hayata},pdfsubject={}, pdfkeywords={}]{hyperref}

\usepackage[dvipdfmx]{graphicx}
\usepackage{amssymb,bm}
\usepackage{amsthm,amsmath}
\usepackage{color}

\usepackage{graphicx}
\usepackage{times}
\usepackage{amssymb,bm}
\usepackage{amsthm,amsmath}

\begin{document}
\preprint{RIKEN-QHP-56}
\title{Time-dependent heavy-quark potential at finite temperature from gauge-gravity duality}

\author{Tomoya Hayata}
\email[]{hayata@riken.jp}
\affiliation{Department of Physics, The University of Tokyo, Tokyo 113-0031, Japan}
\affiliation{Theoretical Research Division, Nishina Center, RIKEN, Wako 351-0198, Japan}
\author{Kanabu Nawa}
\email[]{knawa@riken.jp}
\affiliation{Theoretical Research Division, Nishina Center, RIKEN, Wako 351-0198, Japan}
\author{Tetsuo Hatsuda}
\email[]{thatsuda@riken.jp}
\affiliation{Theoretical Research Division, Nishina Center, RIKEN, Wako 351-0198, Japan}
\affiliation{IPMU, The University of Tokyo, Kashiwa 277-8583, Japan}

\date{September 10, 2013}

\begin{abstract}

 The potential between  a heavy quark and an antiquark inside the quark-gluon plasma is studied
 on the basis of the gauge-gravity duality. 
 A real-time complex potential $V_{Q\bar{Q}}(t,r)$ is derived from the Wilson loop
 which is evaluated by its gravity dual in the Euclidean five-dimensional anti-de-Sitter black hole metric. 
 To make the analytic continuation from the imaginary time to the real time, 
 specific variational configurations of the string world sheet in the Euclidean metric are introduced. 
 A rapid approach of $V_{Q\bar{Q}}(t,r)$ to its stationary value is found at the time scale 
 $t \simeq (\pi T)^{-1}$, and the imaginary part of $V_{Q\bar{Q}}(\infty,r)$ becomes significant 
 above the length scale $r = 1.72 (\pi T)^{-1}$.
 Also, these scales are independent of the 't Hooft coupling $\lambda$.
 Implications of these results to the properties of heavy quarkonia in quark-gluon plasma are briefly discussed.

 \end{abstract}

\maketitle

 The quark-gluon plasma (QGP), whose properties are important
 to understand the physics of early Universe at $10^{-4}\sim10^{-5}$ sec after the big bang, 
 is actively studied in heavy-ion collision experiments
 at the Relativistic Heavy Ion Collider (RHIC) and the Large Hadron Collider (LHC). 
 The experimental data suggest that QGP with the temperature $(T)$ of a few hundred MeV 
 is not a weakly interacting gas of quarks and gluons 
 but rather a strongly coupled quark-gluon plasma (sQGP)~\cite{Sarkar:2010zza}. 
 One of the key quantities which characterizes sQGP is a small ratio of shear viscosity to entropy density
 $\eta/s = (4\pi)^{-1}$~\cite{Kovtun:2004de} obtained 
 by using the holographic duality between $\mathcal{N}=4$ super-Yang-Mills theory and
 the classical supergravity on $\mathrm{AdS}_{5}\times \mathrm{S}^5$~\cite{Maldacena1998I}.

 Another quantity which can probe sQGP is the spectra of heavy quarkonia  at finite $T$~\cite{leptons}. 
 It was originally proposed in the context of the string breaking and Debye screening at finite $T$~\cite{Matsui1986,Satz:2005hx}. 
 Later, the lattice quantum chromodynamics (QCD) simulations~\cite{Asakawa2004} and the QCD sum rule analyses~\cite{Gubler}
 have been attempted to study the melting pattern of quarkonia. 
 It was also realized that the heavy-quark potential at finite $T$ can be defined from the {\it Minkowski} Wilson loop in the thermal medium:
 The potential thus defined is found to be {\it time-dependent} and {\it complex} 
 by using thermal perturbation theory~\cite{Laine2007} and lattice QCD simulations~\cite{Rothkopf2012}. 
 (See also Ref.~\cite{akamatsu}.)

 The purpose of this paper is to derive such a time-dependent complex potential in sQGP
 on the basis of the gauge-gravity duality in the large $N_{\mathrm{c}}$
 and large 't Hooft coupling $\lambda=g_{\rm YM}^2 N_{\rm c}$. 
 We start with an {\it Euclidean} Wilson loop $W_{\rm E}(\tau,r)$ ($\tau$ ($r$)
 is the temporal (spatial) extent with $0\leq\tau < \beta=T^{-1}$). 
 After evaluating $W_{\rm E}(\tau,r)$ by its gravity dual in the {\it Euclidean} AdS$_5$ black hole metric 
 following Maldacena's conjecture~\cite{Maldacena1998II}, 
 we carry out an analytic continuation to obtain the {\it Minkowski} Wilson loop $W_{\rm M}(t,r)$ with real-time $t$. 
 Then, we extract the heavy-quark potential $V_{Q\bar{Q}}(t,r)$ from $W_{\rm M}(t,r)$. 
 We note that there is a previous attempt to derive a complex potential
 under the {\it Minkowski} AdS$_5$ black hole metric~\cite{Kovchekov2008}: 
 Its relation to our approach and its limitation in studying the time-dependent phenomena 
 will also be discussed.

 Let us start with the rectangular Wilson loop $W_{\mathrm{E}}(\tau,r)$
 in the Euclidean space-time and its spectral decomposition~\cite{Rothkopf2012}:
\begin{equation}
  W_{\mathrm{E}}(\tau,r)=\lim_{M\rightarrow \infty}
  \int_{-2M}^{\infty}\mathrm{d}\omega\;\mathrm{e}^{-\omega \tau}\rho(\omega,r).
  \label{spectral1}
\end{equation} 
 Here $M$ is a bare heavy-quark mass taken to infinity at the end,
 and $\rho(\omega,r)$ is the spectral function of the heavy $Q\bar{Q}$ with relative distance $r$ in thermal environment.
 The frequency $\omega$ denotes energy relative to $2M$.
 The heavy-quark potential $V_{Q\bar{Q}}(t,r)$ at finite $T$
 in the Minkowski metric
 is obtained from the analytic continuation of 
 Eq. (\ref{spectral1}), $W_{\mathrm{M}}(t,r)=W_{\mathrm{E}}(\tau\rightarrow it ,r)$~\cite{Laine2007,Rothkopf2012}:
\begin{eqnarray}
  V_{Q\bar{Q}}(t,r)
 = i\partial_t\ln W_{\mathrm{M}}(t,r) 
 =  \frac{\int_{-\infty}^{\infty}\mathrm{d}\omega\;
    \mathrm{e}^{-i\omega t}\omega\rho(\omega,r)}{\int_{-\infty}^{\infty}\mathrm{d}\omega\;\mathrm{e}^{-i\omega t}\rho(\omega,r)}.
  \label{potential1}
\end{eqnarray}
 Although $\rho(\omega,r)$ and $W_{\mathrm{E}}(\tau,r)$ are both real and positive,
 $W_{\mathrm{M}}(t,r)$ and $V_{Q\bar{Q}}(t,r)$ become complex after the analytic continuation. 
 The major questions to be addressed in this paper are 
 (i) how fast the real-time potential approaches its asymptotic value as a function of $t$ in sQGP and 
 (ii) what the typical time-scale and length-scale characterizing the imaginary part of the potential would be.

 Following ~\cite{Maldacena1998II}, 
 we consider the gravity dual description of $W_{\rm E}(\tau,r)$ given by 
 the extremum of the Nambu-Goto (NG) action in the background metric of the classical supergravity: 
\begin{equation}
  W_{\rm E}(\tau,r)=\mathrm{e}^{-S_{\mathrm{NG}}(\tau,r,X^{\mu})} ,
  \label{eq:WES}
\end{equation}
 where $X^{\mu}$ $(\mu=0,\ldots,4)$
 are the coordinates of string world sheet embedded in the five-dimensional Euclidean space-time 
 with a finite rectangular contour on the boundary. 
 These coordinates satisfy the equation of motion, $\delta S_{\mathrm{NG}}/\delta X^{\mu}=0$. 
 We adopt the Euclidean AdS$_5$ black hole metric as a dual description
 for a strongly coupled large $N_{\mathrm{c}}$ gauge theory at finite $T$: 
\begin{equation}
  \mathrm{d}s^2=\ell_s^2\Bigl[\frac{u^2}{R^2}\Bigl\{f(u)\mathrm{d}{\tau^{\prime}}^2+\mathrm{d}\bm{x}^2\Bigr\}
    +\frac{R^2}{u^2}\frac{\mathrm{d}u^2}{f(u)}\Bigr] ,
\end{equation}
 where $\ell_s$ is the string length and $f(u)=1-u_h^4/u^4$ 
 with $u_h=\pi TR^2$ being the location of the event horizon in the fifth coordinate. 
 $R=(2\lambda)^{1/4}$ is a radius in the string unit.

\begin{figure}
\includegraphics[scale=0.16]{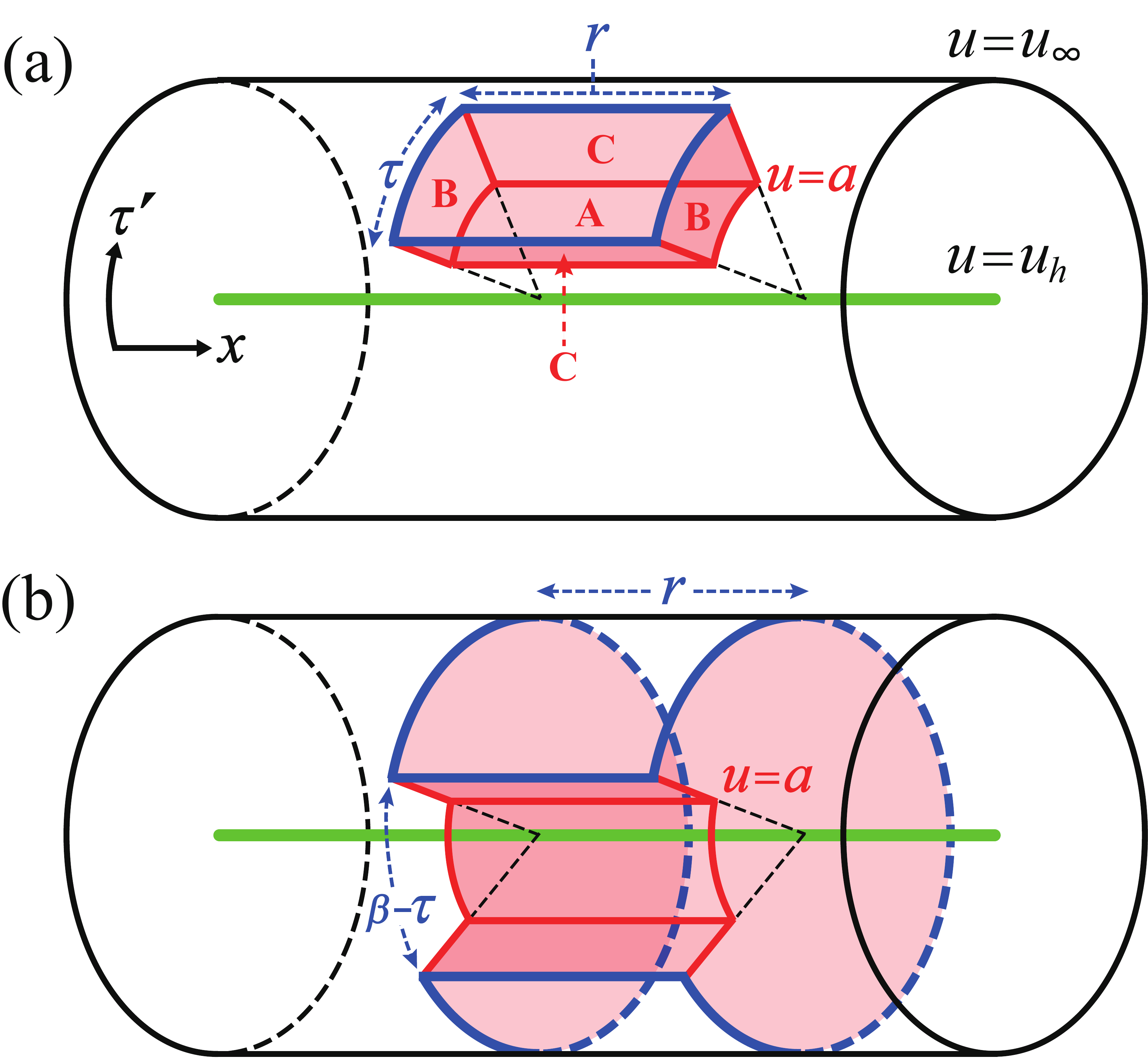}%
\caption{\label{fig1}
Configuration ansatz of world sheet
in the AdS$_5$ black hole metric with a rectangular contour on the boundary. 
 Configurations I, II, and III correspond to (a) with $a>u_h$, 
(a) or (b) with $a=u_h$, and (b) with $a>u_h$, respectively.
}
\end{figure}
\begin{figure}
\includegraphics[scale=.60]{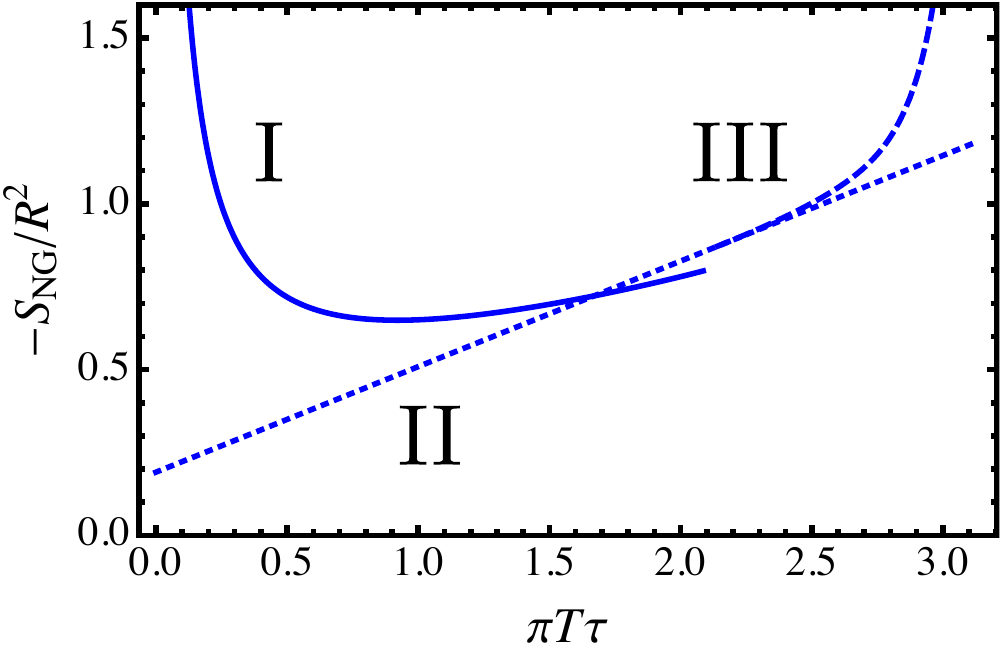}%
\caption{\label{fig2}
Wilson loop as a function of the Euclidean time $\tau$ for $r=(\pi T)^{-1}$. The solid, the dotted and the dashed lines correspond to the configurations I, II and III, respectively.
}
\end{figure}

 It is a formidable task to solve analytically the partial differential equation
 for string world sheet with finite rectangular contour on its boundary. 
 Even if one could solve the problem numerically,
 it is not trivial to make an analytic continuation $\tau \rightarrow it$  only with the numerical data of $W_{\rm E}(\tau,r)$.
 In this paper, we take a variational approach to solve the string world sheet within specific configurations, 
 so that one can pick up qualitative features of the Euclidean solution as well as its analytic continuation. 
 Following the definition of the NG action in Eq.~(\ref{eq:WES})
 as the area of the string world sheet multiplied by the string tension, 
 we evaluate the area of our variational configurations.

 For small $\tau$, we consider a variational configuration I (box type)
 shown in Fig.~\ref{fig1}(a): It has only one variational parameter, 
 the bottom stringy coordinate $a$ in the fifth direction. 
 The treatment of the  cusp at each corner of the Wilson loop  will be discussed later.
 As $\tau$ grows, another configuration II (wedge type) having $a=u_h$ would be relevant. 
 For $\tau$ getting close to $\beta$,
 we consider the variational configuration III (antibox type) shown in Fig.~\ref{fig1}(b)~\cite{Nawa2013}.

 By substituting the configuration I into the NG action, we obtain a formula
 with a single variational parameter $a$:
\begin{eqnarray}
  & & S_{\mathrm{NG}}^{\mathrm{(I)}}(\tau,r,a)=\frac{1}{2\pi}\Bigl[\tau r\frac{a^2}{R^2}\sqrt{f(a)}
    +2\tau (u_{\infty}-a) \notag \\
    & &\ \ \ \ \ \ \ \ \ 
    + 2r\Bigl(\int_{a}^{u_{\infty}}\mathrm{d}u\;\frac{1}{\sqrt{f(u)}}\Bigr) -2 (\tau+r)
    u_{\infty} \Bigr] 
  \label{boxNG0} \\
  & &\ \ 
  =\frac{1}{2\pi}\Bigl[\tau r\frac{a^2}{R^2}\sqrt{f(a)}-2\tau a-2raF(-\frac{1}{4},\frac{1}{2},\frac{3}{4};\frac{u_h^4}{a^4})\Bigr],
  \label{boxNG}
\end{eqnarray}
 with $F$ being the Gauss hypergeometric function. 
 The first, second and third terms of Eq.~(\ref{boxNG0}) come from the area A, B and C in Fig.~\ref{fig1}(a), respectively. 
 The forth term in Eq.~(\ref{boxNG0}) corresponds to the subtraction of
 the quark self-energy $2(\tau+r)(2\pi)^{-1}u_{\infty}$
 in the the temporal and spatial direction~\cite{Maldacena1998II}: 
 The action becomes finite for $u_{\infty}\rightarrow\infty$ after this subtraction. 
 The extremum condition $\delta S^{\mathrm{(I)}}_{\mathrm{NG}}/\delta a=0$ given $\tau$ and $r$ reads
\begin{equation}
  \bar{a}-\frac{1}{\tau}=\frac{1}{r}\sqrt{1- \left( \frac{\pi T}{\bar{a}} \right)^4 } , 
  \label{ex1}
\end{equation}
 with $\bar{a}\equiv aR^{-2}$. 
 The solution of Eq. (\ref{ex1}) obeys the scaling law,
 $\bar{a}(\tau,r,T)=\pi T f(\pi T\tau$, $\pi Tr)$. Therefore, 
 characteristic space-time scale obtained from  $\bar{a}(\tau,r,T)$ would be  $(\pi T)^{-1}$.
 The configuration III gives  similar equation for $a(\tau,r)$ through
 $S_{\mathrm{NG}}^{\mathrm{(III)}}(\tau,r,a)$ ~\cite{Nawa2013}.

 In Fig.~\ref{fig2}, we plot $W_{\rm E}(\tau,r)$ obtained by
 minimizing $S_{\mathrm{NG}}$ in each configuration as a function of $\tau$ with fixed $r$. 
 We consider only the real solution of $a(\tau,r)$ 
 as physically acceptable one in the Euclidean space-time. 
 It turns out that such a real solution disappears for configuration I (III)
 for large (small) $\tau$~\cite{footnote0}. 
 Although the configurations I, II and III are not connected smoothly, 
 either the extended variational configurations or the fluctuations of
 the string world sheet would eventually remove the nonsmoothness. 
 If we consider the branch with larger value of $-S_{\mathrm{NG}}$ given $\tau$ in Fig.~\ref{fig2}, 
 the result is qualitatively consistent with that obtained from lattice QCD simulations~\cite{Rothkopf2012}.

 Within our variational approach, 
 the analytic continuation of $W_{\mathrm{E}}(\tau,r)$ is equivalent to 
 solve Eq.~(\ref{ex1}) with $\tau \rightarrow it$.
 Below,  we focus on the configuration I; it should correctly describe the short-time dynamics
 which is one of the main aims of this paper. 
 Moreover, the potential obtained from configuration I  is qualitatively consistent with the known results
 even in the asymptotic (static) limit as will be shown later. 
 We mention here that the configuration III  may receive sizable correction
 from the  mixing between the thermal Wilson loop and the Polyakov loop correlation~\cite{Rey1998}
 as indicated in thermal perturbation theory~\cite{Berwein2012}.
 In the gauge-gravity duality, such a mixing appears beyond 
 the classical supergravity adopted in this paper.

 We solved Eq. (\ref{ex1}) numerically with $\tau\rightarrow it$
 and found a {\it unique} solution $a(it,r)$
 consistent with $a(\tau \sim 0,r)= \sqrt{2\lambda}\tau^{-1} + \cdots $ for the configuration I. 
 Substituting the solution into Eq.~(\ref{boxNG}), we obtain $W_{\rm M}(t,r)$  from $W_{\rm E}(it,r)$.  
 Then, Eq. (\ref{potential1}) leads to
\begin{eqnarray}
  & & V_{Q\bar{Q}}(t,r)
  \notag \\
  & & \ = -\frac{\sqrt{2\lambda}}{2\pi r}
  \left[ 2r\bar{a}(it,r) - (r \bar{a}(it,r))^3 
    -i\frac{r}{t} (r\bar{a}(it,r))^2 \right].
  \label{potential2}
\end{eqnarray}

 Since $\bar{a}(it,r)$ does not depend on $\lambda$, the potential is proportional to $\sqrt{\lambda}$ irrespective of the values of $T$, $r$ and $t$. 
 Note here that $a(it,r)$ is complex, while $a(\tau,t)$ is real: This is the reason why $V_{Q\bar{Q}}(t,r)$ can become complex while $W_{\mathrm{E}}(\tau,r)$ is real. 
 To check if the above potential is consistent with the known result, we consider the asymptotic limit $t\rightarrow\infty$ at $T=0$ in Eq.~(\ref{potential2}): 
 In this case, $\bar{a}(it,r) \rightarrow r^{-1}$ from Eq.~(\ref{ex1}), so that we obtain 
 $V_{Q\bar{Q}}(\infty,r)\Big|_{T=0}=-\sqrt{2\lambda}(2\pi r)^{-1}$. 
 This is the same as that given in Ref.~\cite{Maldacena1998II}, except that the numerical coefficient $(2\pi)^{-1}$ 
 is smaller by $30\%$ than the exact value $4\pi^2\Gamma(1/4)^{-4}$.

\begin{figure}[t]
\includegraphics[scale=.60]{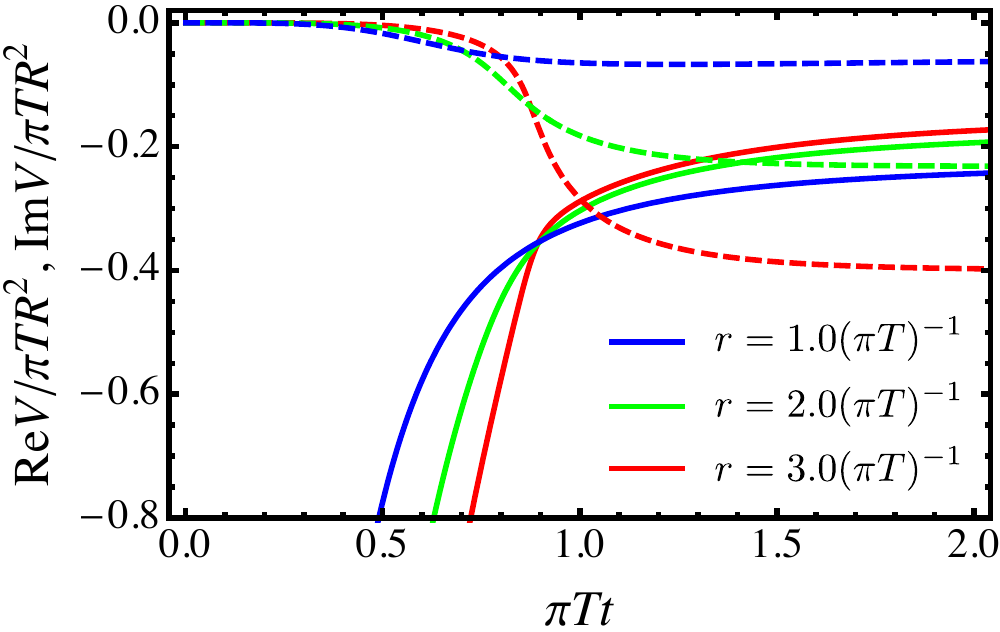}%
\caption{\label{fig3} Real (imaginary) part of the potential denoted by 
the solid (dashed) lines as a function of the real-time $t$ for $r=1.0(\pi T)^{-1}$, $2.0(\pi T)^{-1}$ and $3.0(\pi T)^{-1}$.}
\end{figure}

 The time dependence of the real and imaginary parts of the potential obtained
 from Eq.~(\ref{potential2}) are plotted in Fig.~\ref{fig3} for typical values of $r$. 
 Irrespective of the values of $r$, the potentials reach their asymptotic values quickly
 near the equilibration time $t_{\mathrm{eq}} \simeq (\pi T)^{-1}$. 
 Note that $t_{\mathrm{eq}}$ is independent of $\lambda$, since $\bar{a}$
 and $(2\lambda)^{-1/2}V_{Q\bar{Q}}(t,r)$ are $\lambda$-independent. 
 This is in contrast to the case of weakly coupled QGP (wQGP) at $N_c=3$
 where we have $t_{\mathrm{eq}} \simeq 10m_{\rm D}^{-1} \sim 10(gT)^{-1}$ for $r=(1 \sim 3)(\pi T)^{-1}$ 
 (see Fig.1 of the second reference in Ref.~\cite{Laine2007}).
 Here the Debye mass in wQGP is given by $m_{\rm D}=gT \sqrt{1+N_f/6}$. 
 One finds that $t_{\mathrm{eq}}$ in sQGP is not only parametrically different
 from that in wQGP, but also by an order of magnitude shorter than it. 
 (See Ref.~\cite{Bak:2007fk} for related observation on $m_{\rm D}$.)

 Here we estimate the contribution
 from the cusp singularity of the rectangular Wilson loop 
 to the time dependence of the potential as discussed above.
 Scale-invariant world sheets
 near the four cusps give additional logarithmic divergence in the action:
 $S_{\rm cusp}\sim -4R^2(2\pi)^{-1}F\left(\pi/2\right)\log(L/\varepsilon)$,
 where $(2\pi)^{-1}F(\pi/2)\simeq 0.1$ and $\varepsilon$ is the ultraviolet cutoff~\cite{Drukker1999}. 
 $L$ should be chosen as the smallest length among $t$, $r$, and $(\pi T)^{-1}$:
 They violate the scale invariance around the cusp and serve as an infrared cutoff.
 If $t$ is the smallest length scale relative to others, $L \sim t$ and 
 a power-law behavior, $V_{\rm cusp}\sim 4R^2(2\pi)^{-1}F\left(\pi/2\right)i/t$, is obtained.
 On the other hand, if  $t > (\pi T)^{-1}$, 
 the time derivative of $S_{\rm cusp}$ vanishes irrespective of $r$ ~\cite{footnote3}. 
 Therefore, the cusp contribution does not change the results of our box ansatz
 as long as $t > t_{\rm eq}\simeq (\pi T)^{-1}$~\cite{footnote4}.

 The equilibration time $t_{\mathrm{eq}}$ can be regarded as a time-scale 
 that the sQGP
  relaxes to its equilibrium state under the external disturbance caused by 
 a color-singlet heavy $Q\bar{Q}$ with separation distance $r$ at $t=0$. 
 Taking typical QGP temperature $T=300$ MeV at RHIC and LHC, we have  $t_{\mathrm{eq}}\simeq 0.22$ fm: 
 This is comparable to the formation time of the low-lying heavy quarkonia from the hard process~\cite{Sarkar:2010zza},
 so that the propagation of the heavy quarkonia in sQGP may well be described 
 by using the complex potential in the asymptotic (static) limit.

 It is in order here to comment on the relation 
 between our approach and that with the Minkowski AdS$_5$ black hole metric~\cite{Kovchekov2008}.
 Our method is based on the spectral decomposition of the  Euclidean
 Wilson loop with finite $\tau$ and its analytic continuation to the real-time $t$. 
 This enables us to study the question of the time-scale as in (i) and (ii)
 mentioned after Eq.(\ref{potential1}).
 Such feature is not addressed in Ref.~\cite{Kovchekov2008}
 where the static limit is taken from the beginning.
 Moreover, our Euclidean approach 
 has an advantage to single out a unique physical solution in the
 Minkowski space through the analytic continuation,
 while in the Minkowski approach a careful treatment of
 the real-time boundary conditions is required~\cite{Son2003}.

\begin{figure}
\includegraphics[scale=.6]{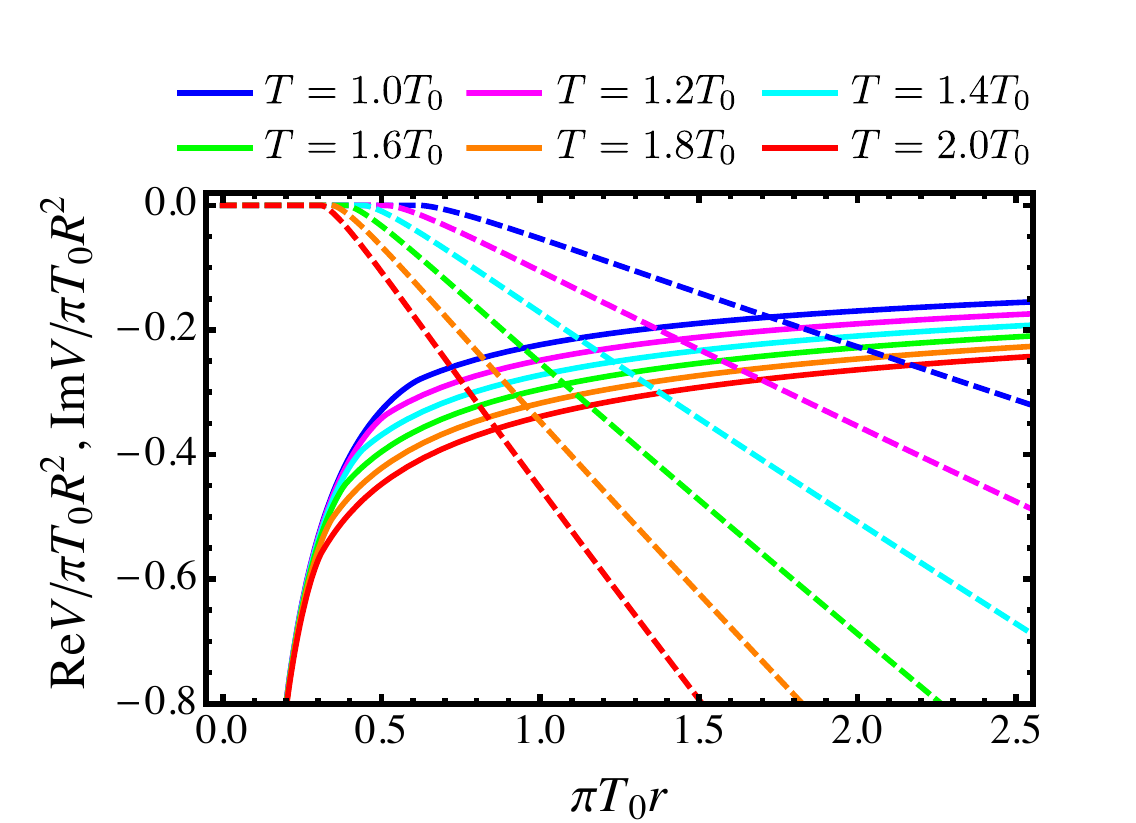}%
\caption{\label{fig4} Real (imaginary) part of the potential in the asymptotic limit ($t \rightarrow \infty$) 
denoted by the solid (dashed) lines as a function of the separation $r$. $T_0$ is an arbitrary energy scale. }
\end{figure}

 In Fig.~\ref{fig4}, the real and imaginary parts of the potential
 after equilibration are plotted for different values of $T$. 
 We normalize the potential by $\pi T_0 R^2$,
 so that the curves are $\lambda$ independent. 
 Here $T_0$ is an arbitrarily energy scale.
 At short distance, ${\rm Re} V_{Q\bar{Q}}(\infty,r)$ shows Coulomb-type behavior irrespective of $T$. 
 On the other hand, at long distance, the potential becomes deeper as $T$ increases; 
 this is consistent with the result in lattice QCD simulations in which the large $r$ behavior is 
 dictated by twice the thermal part of single-quark free energy $2F_Q(T)$~\cite{Maezawa:2011aa}. 
 Note that such a deepening of the real part at long distances can be also seen in Ref.~\cite{Kovchekov2008} 
 if we use the same definition of ${\rm Re} V_{Q\bar{Q}}$ as ours 
 by subtracting only the $T$-independent divergence due to the bare quark mass. 
 We should remark here that ${\rm Re} V_{Q\bar{Q}}$ has a long tail $\sim r^{-1/3}$ at large $r$
 and does not approach a constant. 
 This may be related to the limitation of our simple box-type ansatz.

 We find that the imaginary part of the potential shown in Fig.~\ref{fig4}, 
 ${\rm Im} V_{Q\bar{Q}}(\infty,r)$ emerges at the threshold distance, 
 $r_{\mathrm{th}} = (4/27)^{1/4} ({\pi T})^{-1}$.
 Then, it grows linearly:
 ${\rm Im} V_{Q\bar{Q}}(\infty,r \gg r_{\rm th})\rightarrow -\sqrt{2\lambda}(2\pi)^{-1}(\pi T)^2r$. 
 This is qualitatively consistent with that obtained in Ref.~\cite{Kovchekov2008}. 
 Let us now introduce a new length characterizing the dissociation of heavy quarkonia.
 We define a ``dissociation'' distance $r_{\rm dis}$
 where the imaginary part of the potential starts to dominate over the real part; 
 for large $t$, Eq. (\ref{ex1}) shows that $r \bar{a}$ is a function of $\pi T r$ only. 
 Therefore, the condition 
 ${\rm Re} V_{Q\bar{Q}}(\infty,r_{\rm dis})={\rm Im} V_{Q\bar{Q}}(\infty,r_{\rm dis})$
 leads to a $\lambda$-independent relation, 
 $r_{\rm dis}=1.72 (\pi T)^{-1}$.
 Taking $T=300$ MeV again, we have $r_{\rm dis}\simeq 0.36$ fm. 
 This is smaller than the  $Q$-$\bar{Q}$ distance $0.5$ fm for $J/\Psi$
 and $0.56$ fm for $\Upsilon$($2$S) 
 but is larger than $0.28$ fm for $\Upsilon$($1$S)~\cite{footnote1}. 
 (For the estimate of the spatial sizes of heavy quarkonia, see Table 3 in Ref.~\cite{Satz:2005hx}.) 
 This indicates that the sequential melting of heavy quarkonia seen
 in the relativistic heavy-ion collision experiments 
 may be closely related to the physics of the imaginary $Q\bar{Q}$ potential
 rather than the color Debye screening. 
 Detailed phenomenological studies are, however,
 necessary to make quantitative comparison between the theory and experiments~\cite{Petreczky}.

 In this paper, we have presented the time-dependent complex potential 
 between heavy quarks $ V_{Q\bar{Q}}(t,r)$ in the strongly coupled QGP
 on the basis of the gauge-gravity duality. 
 Our starting point is a rectangular Wilson loop in Euclidean space-time, $W_{\rm E}(\tau,r)$. 
 Associated classical configurations of the string world sheet
 under the Euclidean AdS$_5$ black hole metric 
 were evaluated variationally with a single parameter $a(\tau,r)$ in the fifth dimension. 
 A simple algebraic equation for $a(\tau,r)$ allows us 
 to carry out the analytic continuation of the Euclidean Wilson loop $W_{\rm E}(\tau,r)$ 
 to the real-time Wilson loop $W_{\rm M}(t,r)$ for which the logarithmic derivative 
 with respect to $t$ gives $V_{Q\bar{Q}}(t,r)$. 
 Technically, the potential receives an imaginary part due to the process of analytic continuation.

 Resultant $V_{Q\bar{Q}}(t,r)$ is found to have interesting properties. 
 It shows a characteristic equilibration time $t_{\rm eq} \simeq (\pi T)^{-1}$
 which is by an order of magnitude smaller 
 than the value obtained from the weak-coupling thermal perturbation theory. 
 Also, the imaginary part of the potential starts
 to dominate over the real part at $r_{\rm dis} = 1.72(\pi T)^{-1}$ 
 which is comparable to the $Q$-$\bar{Q}$ distance of the low-lying heavy quarkonia. 
 The above time scale and the length scale indicate the validity of
 using $t$-independent heavy quark potential in sQGP and 
 the relevance of the imaginary part of the potential
 on the sequential melting of heavy quarkonia in sQGP, respectively.

 We have several future directions to improve our analyses presented in this paper. 
 First, we need to extend our variational configurations
 so that a single smooth function for $W_{\rm E}(\tau,r)$ is obtained. 
 It would be also necessary to check the validity of various variational ansatz
 by solving the equation of motion in the Euclidean metric numerically without approximation. 
 Such a numerical solution may be also used to reconstruct the spectral function $\rho(\omega,r)$ 
 by the Bayesian technique.

\begin{acknowledgments}

 The authors thank H.~Ooguri, Y.~Imamura, S.~Nakamura, M.~Natsuume,
 A.~Miwa, Y.~Hidaka, Y.~Hatsuda, Y.~Makeenko and Y.~Kim for stimulating discussions and helpful comments.
 T.~Hayata is supported by JSPS Research Fellowships for Young Scientists.
 K.~Nawa is supported by the Special Postdoctoral Research Program of RIKEN. 
 T.~Hatsuda is supported in part by MEXT Grant-in-Aid for Scientific Research on Innovative Areas (Grant No.2004:20105003) 
 and by JSPS Grant-in-Aid for Scientific Research (B) Grant No.22340052.

\end{acknowledgments}


\bibliography{}

\begin{thebibliography}{100}
\bibitem{footnote0}
 For configuration I, there exist two real solutions, 
 a local minimum (maximum) of $S_{\mathrm{NG}}$.
 For large $\pi T \tau$, they undergo a pair annihilation and
 the real solution disappears. However, such an annihilation 
 happens only after the configuration II takes over the 
 configuration I. As for the configuration III, the real
 solution does not appear for $\pi T \tau < \pi -1$.
 Furthermore, the configurations II and III take the same $S_{\mathrm{NG}}$
 at $\pi T \tau = \pi -1$.
\bibitem{footnote3}
 Similar situation has also been found in the weak-coupling perturbation,
 Eq.(3.10) of Ref.~\cite{Laine2007}:
 $V(t,r)$ behaves as $i/t$ for $t\ll r$, while it is highly suppressed as
  $i r^2/t^3$  for $t\gg r$.
\bibitem{footnote4}
By generalizing the box ansatz near the cusp,
 the potential from the cusp can be suppressed already
  at $t \sim 0.5(\pi T)^{-1}$. This is  because 
the bottom of the world sheet touches the event horizon at $\tau \sim 0.5(\pi T)^{-1}$
and $r\sim 0.5(\pi T)^{-1}$.
Such suppression is expected to occur for even smaller $t$  in the exact configuration 
for which the world sheet tends to hang deeper.
\bibitem{footnote1}
If we identify $r_{\rm dis}$ with the typical $Q$-$\bar{Q}$ distances in the heavy quarkonia, 
we find the dissociation temperature $T_{\rm dis}=390$, $220$ and $190$ MeV for $\Upsilon$($1$S), $J/\Psi$ and $\Upsilon$($2$S), respectively.
\end{thebibliography}

\end{document}